\documentclass[final,5p,times,twocolumn]{elsarticle}

\usepackage{amssymb}

\journal{Physics Letters A}

\begin{document}

\begin{frontmatter}



\title{Controlling the energy gap of graphene by Fermi velocity engineering}

\author{Jonas R. F. Lima}
\address{Instituto de Ciencia de Materiales de Madrid (CSIC) - Cantoblanco, Madrid 28049, Spain}
\ead{jonas.iasd@gmail.com}

\begin{abstract}
The electronic structure of a single-layer graphene with a periodic Fermi velocity modulation is investigated by using an effective Dirac-like Hamiltonian. In a gapless graphene or in a graphene with a constant energy gap the modulation of the Fermi velocity, as expected, only changes the dispersion between energy and moment, turning the minibands narrower or less narrow than in the usual graphene depending on how the Fermi velocity is modulated and the energy gap remains the same. However, with a modulated  energy gap it is possible to control the energy gap of graphene by Fermi velocity engineering. This is based on a very simple idea that has never been reported so far. The results obtained here reveal a new way of controlling the energy gap of graphene, which can be used in the fabrication of graphene-based devices.
\end{abstract}


\end{frontmatter}


\section{Introduction}

Graphene, a one-atom thick layer of carbon atoms arranged in a hexagonal structure,  has attracted a great research interest since its first successful experimental fabrication in 2004 \cite{Novoselov}. Such interest is due, for instance, to its intriguing physical properties and the wide range of potential applications \cite{RevModPhys.81.109,RevModPhys.82.2673,RevModPhys.83.407}. One of the most interesting features of graphene is that its low-energy electronic structure can be described by using a Dirac-type Hamiltonian. As a consequence, due to the fact that there is no energy gap in its electronic structure, the Klein tunneling \cite{Katsnelson,PhysRevB.73.241403} prevents charge carriers in graphene from being confined by an electrostatic potential. Thus, in order to use graphene in electronic devices, different ways of opening and controlling an energy gap in its electronic structure have been investigated and it remains a subject of great interest.

An energy gap can be induced in graphene, for instance, by substrate. A SiC substrate leads to the opening of a gap of $\approx 0.26$ eV in graphene \cite{SiC}, while a hexagonal boron nitride (\textit{h}-BN) substrate induces a gap of the order of 30 meV which depends on the commensurability between the two lattices \cite{guinea1,guinea2}. It is also possible to open and control an energy gap in graphene by doping, for instance, with boron \cite{bdg,bdg2} or nitrogen \cite{ndg} atoms as well as by strain engineering \cite{Pereira,Low,Guinea20121437}. Besides, it is possible to suppress the Klein tunneling creating confined states in graphene with a spatially modulated gap \cite{peres, Giavaras1,Giavaras2,PhysRevB.86.205422}, whereas the band structure of graphene can be engineered by applying an external periodic potential \cite{Ramezani,Sankalpa,Peeters,Barbier}. Even though different ways of controlling the energy gap in graphene have already been reported, in this paper we show the possibility of doing this with a modulated Fermi velocity.

The Fermi velocity plays an important role in the study of a material, since it carries information on a variety of fundamental properties. The potential application of graphene in electronic devices has motivated the examination of its Fermi velocity engineering. In contrast to a Galilean invariant theory such as Fermi Liquids, the Fermi velocity in graphene increases when electron-electron interactions increase, since graphene is described by an effective field theory that is Lorentz invariant \cite{Valeri}. By changing the carrier concentration in graphene the Fermi velocity can reach $\approx 3\times 10^6 $ m/s \cite{Elias}, whereas with weak electron-electron interactions the Fermi velocity is expected to be $0.85 \times 10^6$ m/s \cite{Hwang}. A modulation of the Fermi velocity can be obtained in graphene, for instance, by placing metallic planes close to the graphene sheet, which will turn electron-electron interactions weaker and, consequently, modify the Fermi velocity \cite{Polini,Yuan}. In this way, depending on how the metallic planes are arranged, it is possible to create regions with different Fermi velocity in graphene, thereby forming velocity barriers. A velocity barrier does not suppress the Klein tunneling in a perpendicular incidence of electrons and the transmittance is always equal to unity \cite{peres}. However, in a non-perpendicular incidence the electrons can be confined and the bound states serve as guide modes. The investigation of the transport properties of graphene with velocity barriers was done in Refs. ~[\cite{Polini,Yuan,Vasilopoulos2,Lei,Wang2013191,Wang2013186,Nian}].

In this paper we investigate the electronic structure of a single-layer graphene with a periodic velocity barrier by using an effective Dirac Hamiltonian. In contrast to the previous investigations, where the transport properties were analyzed, we will study the electronic properties and analyze the effects of a modulated Fermi velocity in the electronic band structure of graphene. In the case of a graphene with a constant energy gap, which includes the case of a gapless graphene, the periodic Fermi velocity changes only the shape of the electron and hole minibands. However, considering both an energy gap and a Fermi velocity modulations, it is possible to control the energy gap in graphene by tuning the Fermi velocity, which is an idea that can be used in the fabrication of graphene-based devices.

The paper is organized as follows. In Sec. II we obtain the dispersion relation for a single-layer graphene with a piecewise constant periodic energy gap and Fermi velocity by using an effective Dirac Hamiltonian. We discuss how this system can be realized experimentally. In Sec. III we analyze the electronic band structure of the system. We show that it is possible to control the energy gap in graphene by a modulated Fermi velocity. We explain in which conditions it happens. The paper is summarized and concluded in Sec. IV.

\section{Model}

\begin{figure}[hpt]
\centering
\includegraphics[width=9cm,height=4cm]{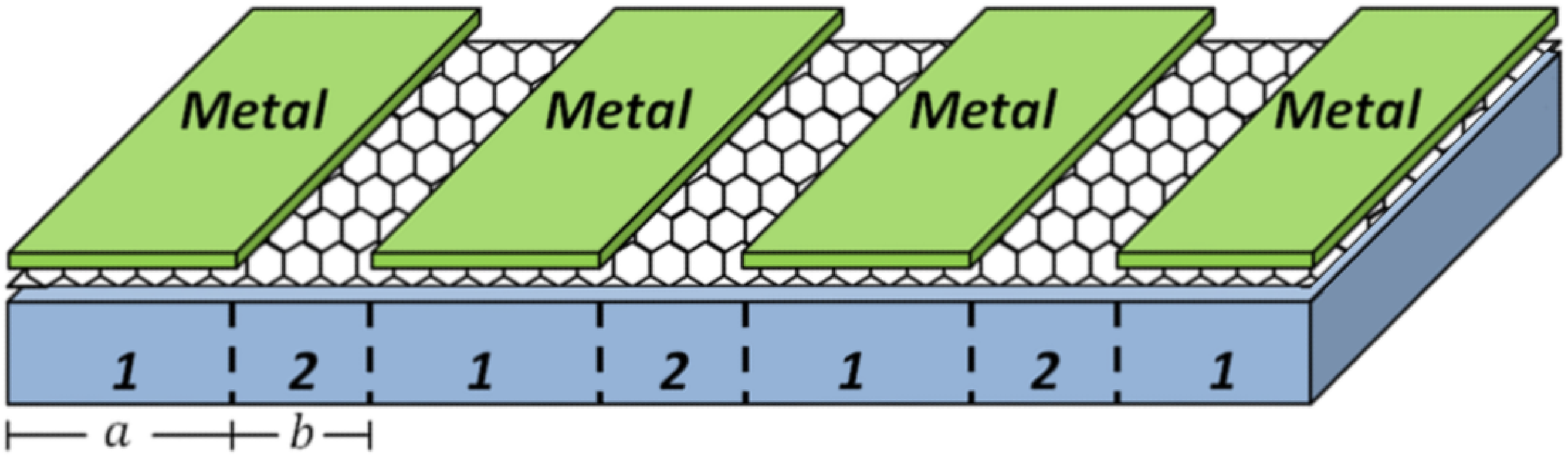}	
\caption{Schematic diagram of a graphene sheet deposited on a heterostructured substrate composed of two different materials which can open different energy gaps in different regions of the graphene sheet inducing a periodic modulation of the energy gap in graphene. Metallic planes are placed close to graphene sheet which induce a periodic velocity barrier. The period of the graphene superlattice is $a+b$.}\label{graphene}
\end{figure}

The effective two-dimensional Dirac Hamiltonian for a single-layer graphene with a position dependent energy gap and Fermi velocity is written as
\begin{eqnarray}
H=-i\hbar \left(\sqrt{v_F(x)}\sigma_x \partial_x  \sqrt{v_F(x)} +v_F(x)\sigma_y \partial_y\right)  \nonumber \\
+\Delta(x)\sigma_z ,
\end{eqnarray}
where $\Delta(x)$ is half of the graphene energy gap, $\sigma_i$ are the Pauli matrices acting on the {\it pseudospin} related to the two graphene sublattices and $v_F(x)$ is the Fermi velocity. The first term on the Hamiltonian above is modified in relation to the usual Dirac operator for graphene in order to have a Hermitian operator \cite{peres}. We are considering that the Fermi velocity and the energy gap change only in the $x$ direction. In Fig. \ref{graphene} there is a schematic diagram of the graphene with a periodic energy gap and Fermi velocity. The graphene is deposited on a heterostructured substrate composed by two different materials, which can open different energy gaps in different regions of the graphene sheet that we will denote by $\Delta_1$ and $\Delta_2$. Metallic planes are placed close to graphene sheet which induce a periodic velocity barrier. The Fermi velocity in each region will be denoted by $v_1$ and $v_2$. The period of the system is $a+b$.

The Dirac equation is given by
\begin{equation}
H\psi(x,y)=E\psi(x,y)
\end{equation}
where $\psi(x,y)$ is a two-component spinor that represents the two graphene sublattices. Writing 
\begin{equation}
\psi(x,y)=e^{-ik_yy}\psi(x)
\end{equation}
and defining $\sqrt{v_F(x)}\psi(x) = \phi(x)$, the Dirac equation becomes
\begin{eqnarray}
-i\hbar v_F(x)\sigma_x\partial_x \phi(x)&+&\left[\Delta(x) \sigma_z \right. \nonumber \\ 
&-&\left. \hbar k_y \sigma_y \right]\phi(x)  =E\phi(x) \; ,
\end{eqnarray}
which can be written as
\begin{equation}
i\frac{d\phi(x)}{dx}=M(x)\phi(x) \; ,
\label{dirac}
\end{equation}
where
\begin{equation}
M(x)=\left(
\begin{array}{cc}
ik_y & \frac{-E-\Delta(x)}{\hbar v_F(x)} \\
\frac{-E+\Delta(x)}{\hbar v_F(x)} & -ik_y
\end{array} \right) \; .
\end{equation}
The solution of Eq. (\ref{dirac}) is given by
\begin{equation}
\phi(x)= \mathcal{P} \exp\left( -i\int_{x_0}^x dx' M(x') \right) \phi(x_0) \; ,
\end{equation}
where $\mathcal{P}$ is the path ordering operator, which places smaller values of $x$ to the right \cite{McKellar,Arovas}. $M(x)$ is a piecewise constant function, so if $x$ and $x_0$ belong to the space-homogeneous region, the solution above can be simplified as
\begin{equation}
\phi(x)=\Lambda(x-x_0) \phi(x_0)  ,
\label{solution}
\end{equation}
where $\Lambda(x-x_0) = \exp[-i(x-x_0)M(x)]$. Expanding this exponential, one can write
\begin{equation}
\Lambda(x-x_0) = \left(
\begin{array}{cc}
\cos\alpha + \frac{k_y}{k(x)}\sin\alpha & -i\sin\alpha \frac{-E-\Delta}{\hbar v_F(x) k(x)}\\
-i\sin\alpha \frac{-E+\Delta}{\hbar v_F(x) k(x)} & \cos\alpha - \frac{k_y}{k(x)}\sin\alpha
\end{array}\right) \; ,
\end{equation}
where $\alpha = (x-x_0)k(x)$ and $k(x)=([E^2-\Delta^2]/\hbar^2 v_F^2 - k_y^2)^{1/2}$. The wave function satisfies the Bloch theorem, so $\psi(a+b)=\exp[ik_x(a+b)]\psi(0)$, where $k_x$ is the Bloch wave vector. From Eq. (\ref{solution}) one can write $\phi(a+b)=\Lambda(a+b) \phi(0)$, which is equivalent to $\psi(a+b)=\Lambda(a+b) \psi(0)$, where $\Lambda(a+b)=\Lambda(b)\Lambda(a)$. Comparing this with the Bloch theorem one can see that 
\begin{equation}
\det[\Lambda(a+b)-\exp[ik_x(a+b)]]=0 \; ,
\end{equation}
which yields the relation $2\cos k_x(a+b)= Tr[\Lambda (a+b)]$. Therefore, the dispersion relation is given by
\begin{eqnarray}
\cos &&(k_x l)=\cos(k_1a)\cos(k_2b) \nonumber \\
&&+\frac{k_y^2\hbar^2 v_1v_2-E^2+\Delta_1 \Delta_2}{\hbar^2 v_1v_2k_1k_2}\sin(k_1a)\sin(k_2b) \; ,
\label{dispersion}
\end{eqnarray}
where $k_1=([E^2-\Delta^2_1]/\hbar^2 v_1^2 - k_y^2)^{1/2}$, $k_2=([E^2-\Delta^2_2]/\hbar^2 v_2^2 - k_y^2)^{1/2}$ and we have defined $l=a+b$. One can note that with $\Delta_1=\Delta_2=0$ and $v_1=v_2$, Eq. (\ref{dispersion}) becomes the liner dispersion relation for a single-layer graphene. The left-hand side of Eq. (\ref{dispersion}) limits the right-hand side to the interval $(-1,1)$. Thus, there will be allowed and forbidden values for the energy, which implies in the appearance of energy bands with gaps.

\section{Electronic Structure}

With the dispersion relation, let us now investigate the electronic structure of the system. In what follows, we will consider a constant period for the graphene superlattice equal to $60$ nm, which is an appropriate value for graphene. We will consider also that $a=b=30$ nm.

\begin{figure}[hpt]
\centering
\includegraphics[width=8.5cm,height=4.1cm]{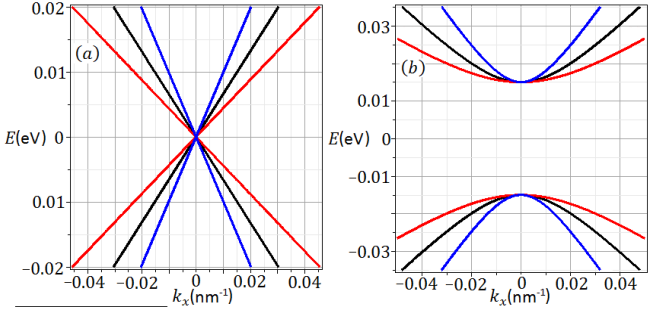}	
\caption{The electron and hole minibands as a function of $k_x$ with $k_y=0$, $\Delta_1=\Delta_2$, $v_1=1\times 10^6$ m/s and $v_2=0.5 \times 10^6$ m/s (red line), $v_2=1\times 10^6$ m/s (black line) and $v_2=3\times 10^6$ m/s (blue line). $(a)$ A gapless graphene. $(b)$ A graphene sheet with an energy gap equal to 30 meV.}
\label{equalgap}
\end{figure} 

In Fig. \ref{equalgap} the dispersion relation (\ref{dispersion}) is plotted for $\Delta_1=\Delta_2$, $v_1=1\times 10^6$ m/s and $v_2=0.5 \times 10^6$ m/s (red line), $v_2=1\times 10^6$ m/s (black line) and $v_2=3\times 10^6$ m/s (blue line). In Fig. \ref{equalgap} $(a)$ we consider a gapless graphene, whereas in Fig. \ref{equalgap} $(b)$ we consider a gapped graphene with an energy gap equal to $30$ meV. One can see that the periodic modulation of the Fermi velocity does not change the energy gap in a graphene sheet with a constant energy gap. In this case, a periodic Fermi velocity modulation only modifies the electron and hole minibands which could be narrower or less narrow than in the usual graphene. Therefore, it is the same of considering a constant Fermi velocity, changing only its value. This explains the fact that the electronic band structure of graphene with a periodic Fermi velocity has not been investigated so far. As it was already mentioned, since several fundamental properties of a material depend on the Fermi velocity, it plays an important role. However, the periodic modulation of the Fermi velocity in this case does not bring any new physics to graphene.

\begin{figure}[hpt]
\centering
\includegraphics[width=8.5cm,height=4.1cm]{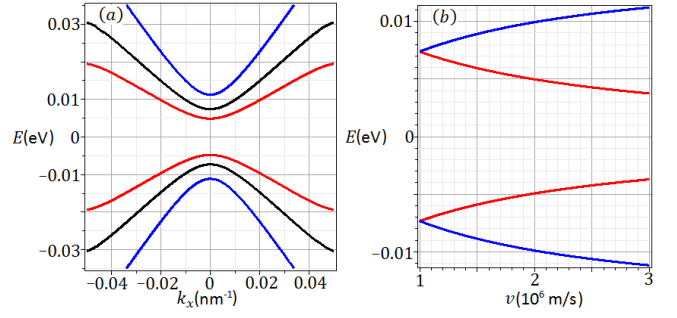}	
\caption{The electron and hole minibands for $\Delta_1=0$ and $\Delta_2=15$ meV. $(a)$ The energy as a function of $k_x$ with $k_y=0$, $v_1=1\times 10^6$ m/s and $v_2=0.5 \times 10^6$ m/s (red line), $v_2=1\times 10^6$ m/s (black line) and $v_2=3\times 10^6$ m/s (blue line). $(b)$ The red lines show how the energy gap changes as a function of $v_1$ with $v_2=1\times 10^6$ m/s, whereas the blue lines reveal the change of the energy gap as a function of $v_2$ at $v_1=1\times 10^6$ m/s. In this case the energy gap of graphene change with the Fermi velocity.}
\label{periodicgap}
\end{figure}

The electron and hole minibands for $\Delta_1=0$ and $\Delta_2=15$ meV are plotted in Fig. \ref{periodicgap}. In Fig. \ref{periodicgap} $(a)$ we consider again $v_1=1\times 10^6$ m/s and $v_2=0.5 \times 10^6$ m/s (red line), $v_2=1\times 10^6$ m/s (black line) and $v_2=3\times 10^6$ m/s (blue line). In this case, in contrast to the previous ones, it is possible to increase or decrease the energy gap in graphene with a periodic modulation of the Fermi velocity. In Fig. \ref{periodicgap} $(b)$ one can see how the electron and hole minibands change with the modulation of the Fermi velocity. The red lines show how the energy gap changes as a function of $v_1$ with $v_2=1\times 10^6$ m/s, whereas the blue lines reveal the change of the energy gap as a function of $v_2$ at $v_1=1\times 10^6$ m/s. It is possible to see that increasing the Fermi velocity in the graphene region with energy gap equal to zero, the energy gap of graphene decreases, whereas if one increases the Fermi velocity in the graphene region with energy gap different from zero, the graphene gap increases. 

Considering that $\Delta_1 \neq \Delta_2$, the energy gap of the graphene superlattice will be a value between $2\Delta_1$ and $2\Delta_2$. It depends on the width of each region in graphene. In the limit when $a\gg b$ the energy gap of graphene goes to $2\Delta_1$, whereas when $a\ll b$ the energy gap of graphene goes to $2\Delta_2$. The periodic modulation of the Fermi velocity works at the same way. Increasing the Fermi velocity in a region of graphene is equivalent to decreasing the width of this region and keeping the Fermi velocity constant. So, in the limit when $v_1 \rightarrow \infty $ the energy gap of graphene goes to $2\Delta_2$, whereas when $v_2 \rightarrow \infty$ the energy gap of graphene goes to $2\Delta_1$. Therefore, it is possible to tune the energy gap of graphene $E_g$ in the range $2\Delta_1\leq E_g \leq 2\Delta_2$.

\section{Conclusions}

In this paper, we report the results of the electronic band structure of a single-layer graphene with a periodic modulation of the energy gap and Fermi velocity by using an effective Dirac equation. The energy gap is included in the Dirac Hamiltonian as a mass term, while it is necessary to modify the Hamiltonian in order to incorporate a position dependent Fermi velocity and have a Hermitian operator. In the case of a gapless graphene or a gapped graphene with a constant energy gap, the periodic modulation of the Fermi velocity works as a constant Fermi velocity turning the electron and hole minibands narrower or less narrow than in the usual graphene depending on the value of the velocity. However, considering a modulation of the energy gap we have shown that it is possible to control the energy gap of the graphene superlattice by a Fermi velocity modulation. We have show in which conditions it happens and in which range it is possible to tune the energy gap in graphene with the Fermi velocity. The results obtained here provide a new way of controlling the energy gap of graphene, which can be used in the fabrication of graphene-based devices.

{\bf Acknowledgements}: This work was partially supported by CNPq and CNPq-MICINN binational.

\bibliographystyle{elsarticle-num} 
\bibliography{ref}

\end{document}